\documentstyle{article}
\begin{document}
\font\fone=cmr10 scaled\magstep3
\font\ftwo=cmr7 scaled\magstep3
\pagestyle{empty}
\vskip 1.5in
\centerline{\fone A Global Uniqueness Theorem}
\centerline{\fone for Stationary Black Holes}
\vskip 0.3in
\centerline{\ftwo G\'abor Etesi}
\vskip 0.1in
\centerline{Institute for Theoretical Physics, E\"otv\"os University}
\centerline{Puskin u. 5-7, Budapest, H-1088 Hungary}
\centerline{{\tt e-mail: etesi@poe.elte.hu}}
\vskip0.3 in
\begin{abstract}
A global uniqueness theorem for stationary black holes is proved as a 
direct consequence of the Topological Censorship Theorem and the topological
classification of compact, simply connected four-manifolds.
\end{abstract}
\vskip 0.3in
\section{Introduction}
There is a remarkable interplay between differential geometry, the theory
of differential equations and the physics of gravitation in the 
famous proof of the uniqueness of stationary black holes. The first proof 
was given in a series of papers by Carter, Hawking, Israel and Robinson
(for a survey see [8], [12]). In the eighties a very elegant shorter proof
was discovered by Mazur [10] who found a hidden symmetry of the 
electromagnetic and gravitational fields. These very deep and difficult
investigations all were devoted to the uniqueness problem of the 
metric on a suitable four-manifold carrying a Lorentzian asymptotically
flat structure in the spirit of Penrose's description of infinity of 
space-times.

More recently physicist's effort is addressed to the {\it topology}
of the event horizons of general (i.e. non-stationary) black holes.
The first theorems were proven by Hawking [8], [9] and later by
Gannon, Galloway [6],[7] and others. Based on the celebrated Topological
Censorship Theorem of Friedman, Schleich and Witt [5] and using energy
conditions  Chru\'sciel and Wald gave a short proof that the event horizon
of a stationary black hole in a ``moment'' is always a sphere.

The question naturally arises what can one say about the topology of the
space-time itself in this case.

On the other hand the final step in the understanding of four-manifolds
making use ``classical'' (i.e. non-physical) methods was done by Freedman
in 1981 who gave a complete (topological) classification of compact simply
connected four-manifolds.
\newpage
\pagestyle{myheadings}
\markright{G. Etesi: Black Hole Uniqueness}

In this paper, referring to the results of Chru\'sciel--Wald and
Freedman, we prove that a {\it global} uniqueness also holds for stationary
black holes and more generally stationary space-times i.e. not only the 
metric but even the topology of the space-time in question is unique. Our
method is based on a natural compactification of the space-time manifold 
and a careful study of a vector field extended to this compact manifold.

Truely speaking, this is not a very surprising result in light of the 
local uniqueness. However, it demonstrates the power of the theorems
mentioned above. 
\section{Vector Fields}
First we define the precise notion of a stationary, asymptotically flat
space-time containing a black hole collecting the standard definitions.
Let us summarize the properties of an asymptotically flat and empty
space-time that we need; for the whole definitions see [12] and the
notion of an asymptotically empty and flat space-time can be found in [8].
\vskip 0.1in
Let $(M, g)$ be a space-time manifold and $x\in M$. then $J^{\pm}(x)$ is
called the causal future and past of $x$ respectively. If a space-time
manifold $(M, g)$ is {\it asymptotically flat and empty} then there exists
a conformal inclusion $i:\:\:(M, g)\mapsto (\tilde{M}, \tilde{g})$ such that
\begin{itemize}

\item $\partial i(M)=\{i_0\}\cup {\cal I}^+\cup {\cal I}^-$, where $i_0$
is the space-like infinity and the future and past null-infinities
${\cal I}^{\pm}$ satisfy ${\cal I}^{\pm}:=\partial J^{\pm}(i_0)\setminus
\{i_0\} = S^2\times {\bf R}$;

\item $\tilde{M}\setminus 
i(M)=\overline{J^+(i_0)}\cup\overline{J^-(i_0)}$;

\item There exists a function $\Omega :\:\: \tilde{M}\mapsto {\bf R}^+$ 
which is smooth everywhere (except possibly at $i_0$) satisfying
$\tilde{g}\vert_{i(M)}=\Omega^2\left( i^{-1}\right)^*g$ and
$\Omega\vert_{\partial i(M)}=0$;

\item Every null geodesic on $(\tilde{M}, \tilde{g})$ has future and past
end-points on ${\cal I}^{\pm}$ respectively.

\end{itemize}
\vskip 0.1in
\hskip 0.21in {\bf Definition.} Let $(M, g)$ be asymptotically flat. $(M, 
g)$ is called
{\it strongly asymptotically predictable} if there exists an open region
$\tilde{V}\subset\tilde{M}$ such that $\tilde{V}\supset\overline{i(M)\cap
J^-({\cal I}^+)}$ and $(\tilde{V}, \tilde{g}\vert_{\tilde{V}})$ is 
globally hyperbolic.
\vskip 0.1in
{\it Remark.} This definition provides that no singularities are visible 
for an observer in $\tilde{M}\cap\tilde{V}$. Moreover one can prove that
$\tilde{M}\cap\tilde{V}$ can be foliated by Cauchy-surfaces $S_t$ 
($t\in {\bf R}$), see [8], [12].
\vskip 0.1in
{\bf Definition.} Let $(M, g)$ be a strongly asymptotically predictable
space-time manifold. If $B:=M\setminus J^-({\cal I}^+)$ is not empty, then
$B$ is called a {\it black hole region} and $H:=\partial B$ is its {\it
event horizon}.
\vskip 0.1in
{\it Remarks.} Moreover we require $(M, g)$ to be {\it stationary} i.e.
there exists a future-directed time-like Killing field $K$ on $(M, g)$.
In this case $H$ is a three-dimensional null-surface in $M$, hence for 
each $t\in {\bf R}$, $H_t:=H\cap S_t$ ($S_t$ is a Cauchy-surface)
is a two-dimensional surface in $M$. We shall assume that $H_t$ (the event
horizon in a ``moment'') is a two dimensional embedded, orientable, 
smooth, compact surface in $M$ without boundary. This is the requirement
the event horizon to be ``regular''. This condition is satisfied by 
physically relevant black hole solutions of Einstein's equations but there
is no {\it a priori} reason to assume it.

Let $(M, g)$ be a maximally extended space-time manifold as above. In the
following considerations we shall focus on {\it one} outer, 
asymptotically flat region of it i.e. a part of $(M, g)$ whose boundary
at infinity in $\tilde{M}$ is {\it connected}. To get such an (incomplete)
manifold we simply cut up $(M, g)$ along one connected component of its 
event horizon. We shall continue to denote this separated part also by
$(M, g)$ ($g$ is the original metric restricted to our domain). We can see
that under the conformal inclusion $i$ the manifold $(M, g)$ has boundary
$\partial i(M)=\tilde{H}\cup{\cal I}^+\cup\{i_0\}\cup{\cal I}^-$ where
$\tilde{H}=i(H)$ and $\tilde{H}$, ${\cal I}^{\pm}$ are {\it connected} now.
\vskip 0.1in
{\bf Proposition 1.} {\it Let $(M, g)$ be a space-time manifold. Then
$K\vert_H\in\Gamma (TH)$.}
\vskip 0.1in
{\it Proof.} Assuming the existence of a point $p\in H$ such that 
$K_p\notin T_pH$, let $\gamma:\:\: (-1, 1)\mapsto M$ be a smooth 
integral curve of $K$ satisfying $\gamma (0)=p$ and $\dot\gamma 
(0)=K_p$. Hence there is an $\varepsilon\in (-1,1)$ such that $\gamma 
(-\varepsilon)\in B$ and $\gamma (\varepsilon)\notin B$. But this means
that 
$\gamma(-\varepsilon)\in J^-({\cal I}^+)$, since it can be connected by an 
integral curve of $K$ to $\gamma (\varepsilon)\in J^-({\cal I}^+)$. Hence 
this assumption led us to a contradiction.
$\Diamond$
\vskip 0.1in
{\bf Corollary.} $H$ is invariant under the flow generated by $K$ on $M$ 
and, being $H$ a null-surface, $K\vert_H$ is a null vector field.
$\Diamond$
\vskip 0.1in
{\it Remarks.} Using a heuristic argument here we can identify $H_t$ up 
to homeomorphism as follows. Since the boundary of $(M, g)$ at infinity 
is homeomorphic to $S^2\times {\bf R}$ we may assume due to the 
stationarity that $H_t$ is homeomorphic to $S^2$. According to recent 
articles one can prove that this is indeed the case for a stationary
black hole [6], [7], [8], [9]. For our purposes it is more important to
refer to a stronger result of Chru\'sciel and Wald [3].

They prove that an outer asymptotically flat region of a stationary
space-time manifold satisfying the null energy condition is simply
connected as a consequence of the Topological Censorship Theorem [5].    
Under suitable additional hypothesis (e.g. the compactness of $H_t$)
it follows that $H_t$ is homeomorphic to a sphere.

Hence, with the aid of these results we get that $H$ is homeomorphic to
$S^2\times {\bf R}$.
\vskip 0.1in
Now let us study the behaviour of the Killing field near the infinity!
Let $\tilde{K}:=i_*K$ induced by the inclusion $i$.
\vskip 0.1in
{\bf Proposition 2.} {\it $\tilde{K}$ becomes a null vector field at 
infinity.}
\vskip 0.1in
{\it Proof.} Let $\tilde{\gamma}:\:\: {\bf R}\mapsto\tilde{M}$ be an 
inextendible integral curve of $\tilde{K}$! We may write:
$$\Vert\tilde{K}_{\tilde{\gamma}
(t)}\vert_{i(M)}\Vert^2=\tilde{g}\vert_{i(M)}
(\tilde{K}_
{\tilde{\gamma} (t)},\:\tilde{K}_{\tilde{\gamma} (t)})=
\Omega^2(\tilde{\gamma} (t))\left( i^{-1}\right)^*g(i_*K_{\gamma
(t)},\:i_*K_{\gamma
(t)})=$$
$$=\Omega^2(\tilde{\gamma} (t))g(K_{\gamma (t)},\:K_{\gamma (t)}).$$
We have used the third property of asymptotic flatness. However, using the
fact that $K$ is a Killing field on $M$, we can write 
$$\Vert \tilde{K}_{\tilde{\gamma} (t)}\vert_{i(M)}\Vert ^2=a\Omega
^2(\tilde{\gamma} 
(t)),$$
where $a:=g(K_{\gamma (t_0)},\:K_{\gamma (t_0)})$ is a constant for 
an arbitrary $t_0\in {\bf R}$. But 
$$ \lim\limits_{t\rightarrow\pm\infty}\Omega ^2(\tilde{\gamma} (t))=0$$
because of the asymptotic flatness.
$\Diamond$
\vskip 0.1in
In the light of Proposition 1. and 2. we can see that $K$ approaches a null
vector field near the boundary of $M$. Hence it is straightforward to
study the behaviour of null vector fields on $M$ and $\tilde{M}$.
Applying a smooth deformation to $K$ on $(M, g)$ we can produce a smooth,
nowhere vanishing (but highly non-unique!) {\it null} vector field $K_0$
on $(M, g)$ whose integral curves are inextendible geodesics. Denoting 
by $\tilde{K_0}$ the image of this field under $i$, i.e. $\tilde{K_0}=
i_*K_0$, $\tilde{K_0}$ has future and past end-points on ${\cal I}^{\pm}$
respectively by the fourth property of asymptotically flat space-times.

{\it Remark.} Of course, we could have started with this null vector 
field instead of the Killing field $K$. The reason for dealing with the 
naturally given Killing field was the attempt to exploit as much as 
possible the structure of a stationary, asymptotically flat space-time 
manifold.

Now let $\tilde{X}$ be an inextendible  null vector field on 
$(\tilde{M}, \tilde{g})$ i.e. its integral curves are inextendible. We 
would like to study the extension of this field to the null infinities
hence first we have to extend the domain of its integral curves, which 
is ${\bf R}$ in this moment. Let us suppose that this extended domain is
the circle $S^1$.
\vskip 0.1in
{\bf Proposition 3.} {\it Let $\tilde{X}$ be a null vector field on 
$(\tilde{M}, \tilde{g})$ with extended domain whose integral curves 
are geodesics. Then $\tilde{X}$ can be extended to ${\cal I}^{\pm}$ if 
and only if $\tilde{X}\vert_{{\cal I}^{\pm}}=0.$}
\vskip 0.1in
{\it Proof.} Let $i(S)\cup\{i_0\}=:\tilde{S}\subset\tilde{M}$ be a 
space-like hypersurface and let $x\in\tilde{S}\setminus 
(\tilde{S}\cap\tilde{B})$ ($x\not=i_0$). Then there exists an integral
curve $\tilde{\gamma} :\: {\bf R}\mapsto\tilde{M}$ such that 
$\tilde{\gamma} (0)=x$. So we can find a point $q\in {\cal I}^+$ 
possessing the property $$q=\lim\limits_{t\rightarrow 
+\infty}\tilde{\gamma} (t).$$
Let us define a map $\phi : \tilde{S}\setminus (\tilde{S}\cap\tilde{B})
\mapsto {\cal I}^+$ by $\phi (x)=q$.      

Let we assume that we have extended $\tilde{X}$ to ${\cal I}^+$ in a 
smooth manner and there is a $q\in {\cal I}^+$ such that $\tilde{X}_q\not 
=0$! Hence there is a smooth curve $\tilde{\beta}:\:(-\varepsilon,
\varepsilon)
\mapsto {\cal I}^+$ for a suitable small positive number $\varepsilon$ 
satisfying
$$\tilde{\beta}(0)=q,\:\:\:\dot{\tilde{\beta}}(0)=\tilde{X}_q.$$   
Obviously we can find a $\delta <\varepsilon$ such that for $q\not=q'\in
U_q$
$$\tilde{\beta}(\delta)=q',\:\:\:\dot{\tilde{\beta}}(\delta)=\tilde{X}_{q'}
\not=0.$$
But in this case there is an $x'\in U_x$ ($U_x$ denotes a small 
neighbourhood of $x$) and an integral curve $\tilde{\gamma}'$ of $\tilde{X}$
such that $\phi (x')=q'$.

In other words $q'$ satisfies
$$q'\in {\bf im}\tilde{\beta},\:\:\:q'\in {\bf im}\tilde{\gamma}',$$
and
$$q\in {\bf im}\tilde{\beta},\:\:\:q\notin {\bf im}\tilde{\gamma}'.$$
This means that $q'$ is a branching point of an integral curve of 
$\tilde{X}$. But in this case even if $\tilde{X}_{q'}$ is well defined
$-\tilde{X}_{q'}$ is not and this is a contradiction.

Similar argument holds for ${\cal I}^-$.
$\Diamond$
\vskip 0.1in
Hence we are naturally forced to find a null vector field  tending to 
zero on the null infinities. However note that on $(\tilde{M}, \tilde{g})$
there is a natural cut-off function, namely $\Omega$. Certainly there 
exists a $k\in {\bf N}$ such that the vector field defined by
$$\tilde{X}_0:=\Omega ^k\tilde{K}_0$$ 
is a zero vector field restricted to ${\cal I}^{\pm}$ because of the third
property of asymptotic flatness. Note that this new vector field can be
extended as zero to $i_0$ as well: Surrounding $i_0$ by a small 
neighbourhood $U$ one can see (since $U\cap ({\cal I}^+\cup {\cal 
I}^-)\not=\emptyset$) $\tilde{X}_0$ is arbitrary small in $U$.   
 
It is straightforward that $\tilde{X}_0$ is not transversal to ${\cal I}^\pm$
in the sense of [1] since it approaches the null-infinities as a 
tangential field. This would cause some difficulties later on in our
construction. Fortunately one can overcome this non-transversality phenomenon
by a general method. Due to standard transversality arguments [1] 
applying a {\it generic} small perturbation to $\tilde{X}_0$ in a suitable
neighbourhood of ${\cal I}^\pm$ we can achieve that the perturbed field
$\tilde{X}_\varepsilon$ will be transversal to the submanifolds of 
null-infinities (and even remains zero on them, of course).
\section{The Compactification Procedure}
Now let $N_1$ be a smooth four-manifold. We call a subset $C\subset N_1$
{\it a domain} of $N_1$ if it is diffeomorphic to a closed four-ball 
$B^4$. Let $V$, $U^\pm\subset N_1$ be domains and $j:\tilde{M}\mapsto N_1$
a smooth embedding satisfying the following conditions:
\begin{itemize}

\item $N_1\setminus j(\tilde{M})={\bf int}V$;

\item There exists a point $p_0\in N_1$ and domains $U^\pm$ satisfying 
$U^+\cap U^-=\{p_0\}$ such that $j(i_0)=p_0$ and
$j({\cal I}^\pm)\subset\partial U^\pm$ and 
$j(\overline{J^\pm (i_0)})\subset U^\pm$;

\item $\partial U^\pm\setminus j({\cal I}^\pm) =: A^\pm\subset\partial V$;

\item $j(\tilde{H})\subset\partial V$;

\item $j(\tilde{H})\cap A^\pm=\emptyset$.

\end{itemize}

It is not difficult to see that such a $j$ exists due to the topology of an
outer region of an asymptotically flat stationary space-time containing a 
black hole and $N_1$ has no boundaries. Consider the vector field
$Y_1:=j_*\tilde{X}_\varepsilon$. We wish to extend this field into the
interior
of $V$.

To be explicit take a coordinate map from $N_1$ to ${\bf R}^4$ under which
$V$ maps to the cylinder $V:=B_1^3\times [-1,1]$ where $B_r^3$ denotes a 
closed three-ball of radius $r$ originated at the origin. Moreover let
$W:=D_2^3\times (-1,1)$ be another {\it open} neighbourhood (where $D_r^3$
denotes the open ball respectively).

We assume that $j(\tilde{H})$ is given by the ``belt'' 
$\partial B_1^3\times (-{1\over 2},{1\over 2})$ and $A^\pm$ are represented
by the top- and bottom-balls $B_1^3\times\{\pm 1\}$. Note that this picture
is consistent with the conditions for $j$ (except that $V$ is not a four-ball
but a cylinder).

Note that by Proposition 1
$$Y_1\vert_{j(\tilde{H})}\in\Gamma(Tj(\tilde{H})).$$
In other words $Y_1$ has the form (0,0,0,1) on $j(\tilde{H})$. Hence we 
define a smooth vector field $Z$ in $W$ as follows: Let $q\in D_2^3$ and
\begin{itemize}

\item for $t\in [-1,-{1\over2}]$
$$Z_{(q,t)}:=(0,0,0,f_-(t));$$
where for the smooth cut-off function $f_-$ the following holds:

\[f_-(t)=\left\{\begin{array}{ll}
0 & \mbox{if $t\leq -1$}\\
1 & \mbox{if $t\geq -{1\over2};$}
\end{array}
\right. \]

\item for $t\in(-{1\over2},{1\over2})$
$$Z_{(q,t)}:=(0,0,0,1);$$

\item for $t\in [{1\over2},1]$
$$Z_{(q,t)}:=(0,0,0,f_+(t));$$
where

\[f_+(t)=\left\{\begin{array}{ll}
1 & \mbox{if $t\leq {1\over 2}$}\\
0 & \mbox{if $t\geq 1.$}
\end{array}
\right.\]

\end{itemize}
Clearly $Z$ is a smooth vector field in $W$ and $Z\vert_{A^\pm}=0$. Now take
a smooth cut-off function $\rho:{\bf R}^4\mapsto{\bf R}^+$ satisfying 
$\rho\vert_V=1$ and being zero on the complement of $W$. Define the extension
of $Y_1$ by
$$\rho Z+(1-\rho)Y_1.$$
It is obvious that the extended field (also denoted by $Y_1$) is a smooth 
vector field on $N_1$ due to the transversality of the original field to
${\cal I}^\pm$ (more precisely it has not been defined in $U^\pm$ yet) and
satisfies $Y_1\vert_{\partial U^\pm}=0$.

As a final step let us apply a smooth homotopy for $N_1$ contracting the
four-balls $U^\pm$ to the point $p_0$. In this way we get a smooth compact
manifold $N_0$ without boundaries and, due to the transversality conditions
on $Y_1$, a well-defined smooth vector field $Y_0$ on it. It is clear that
$Y_0$ has only one (degenerated) isolated singular point, namely $p_0$. Its
index is +2 as easy to see due to the fourth property of asymptotic flatness.
\vskip 0.1in
{\bf Theorem.} {\it $N_0$ is homeomorphic to the four-sphere $S^4$.}
\vskip 0.1in
{\it Proof.} First it is not difficult to see that $N_0$ is simply connected.
Note that $N_0$ is a union of an outer region of the original stationary,
asymptotically flat space-time $M$ and a solid torus-like space $T$ 
homeomorphic to $B^3\times S^1$ with one $B^3$ pinched into a point (namely
which corresponds to $p_0$). Choosing $p_0$ as a base point consider the loop
$l$ in $N_0$ representing the generator of the fundamental group 
$\pi_1(T)={\bf Z}$. Clearly this loop can be deformed continuously into
$M\cup\{p_0\}.$ But referring again to the theorem of Chru\'sciel and 
Wald [3]
the deformed loop $l$ is homotopically trivial since $M$ is simply connected.
It is obvious that every other loops in $N_0$ are contractible proving
$\pi_1(N_0)=0$.

Secondly, we have constructed a smooth vector field $Y_0$ on $N_0$ having one
isolated singular point $p_0$ with index +2. Taking into account that $N_0$
is a smooth, compact manifold without boundaries this means that
$$\chi(N_0)=2$$
according to the classical Poincar\'e--Hopf Theorem [2] [10].

The Euler characteristic of a simply connected compact four-manifold $S$ 
always has the form $\chi(S)=2+b_2$ where $b_2$ denotes the rank of its 
intersection form (or second Betti number). In our case this rank is zero,
hence our form is of even type. By the uniqueness of the trivial zero rank
matrix representing this even form and referring to the deep
theorem of Freedman [4] which gives a full classification of simply connected
compact topological four-manifolds in terms of their intersection matrices
we deduce that $N_0$ is homeomorphic to the four-sphere since its even 
intersection form is given by the same zero-matrix. $\Diamond$
\vskip 0.1in
{\bf Corollary.} The uniqueness of $N_0$ implies the uniqueness of the 
original space-time manifold (more precisely one connected piece of its 
outer region) $M$ since we simply have to remove a singular solid torus
$T$ from $N_0$ and this can be done in a unique way being $N_0$ simply 
connected. Hence, taking into account the explicit example of 
the Kerr-solution, this outer region is always homeomorphic to 
$S^2\times {\bf R}^2.$ $\Diamond$ 
\section{Conclusion}
We have proved that topological uniqueness holds for space-times carrying
a stationary black hole. Note that we could prove only a topological 
equivalence although our constructed manifold $N_0$ carries a smooth 
structure, too. It would be interesting to know if this smooth 
structure was identical to the standard one in light of the 
unsolved problem of 
the four dimensional Poincar\'e-conjecture in the smooth category.

From the physical point of view this uniqueness is important if we are 
interested in problems concerning the whole structure of such space-time
manifolds.

For example one may deal with the description of the vacuum structure of
Yang-Mills fields on the background of a gravitational configuration 
containing a black hole or some other singularity (taking the singularity
theorems into consideration this question is very general and natural). 
One would expect that a black hole may have a strong influence on these 
structures and using our theorem presented here we can study this problem
effectively in our following paper.

Moreover we hope that our construction works for non-stationary i.e. higher
genus black holes as well giving an insight into the structure of such more 
general space-time manifolds.
\section{Acknowledegment}
The work was partially supported by the Hungarian Ministry of Culture and
Education (FKFP 0125/1997).
\section{References}
\hskip 0.21in 
1 V.I. Arnold: {\it Geometrical Methods in the Theory of Ordinary 
Differential Equations}, Springer-Verlag (1988);

2 R. Bott, L. Tu: {\it Differential Forms in Algebraic Topology},
Springer-Verlag (1982);

3 P.T. Chru\'sciel, R.M. Wald: {\it On the Topology of Stationary Black
Holes}, Class. Quant. Grav. {\bf 11}, 147-152 (1994);

4 M.H. Freedman: {\it The Topology of Four-Manifolds}, Journ. Diff. 
Geom. {\bf 17}, 357-454 (1982);

5 J.L. Friedman, K. Schleich, D.M. Witt: {\it Topological Censorship},
Phys. Rev. Lett. {\bf 71}, 1486-1489 (1993);

6 G. Galloway: {\it On the Topology of Black Holes}, Commun. Math. Phys.
{\bf 151}, 53-66 (1993);

7 D. Gannon: {\it On the Topology of Space-like Hypersurfaces, Singularities
and Black Holes}, Gen. Rel. Grav. {\bf 7}, 219-232 (1976);

8 S.W. Hawking, G.F.R. Ellis: {\it The Large Scale Structure of 
Space-Time}, Cambridge Univ. Press. (1973);

9 S.W. Hawking: {\it Black Holes in General Relativity}, Commun. Math. 
Phys. {\bf 25}, 152-166 (1972);

10 P.O. Mazur: {\it Black Hole Uniqueness from a Hidden Symmetry of 
Einstein's Gravity}, Gen. Rel. Grav. {\bf 16}, 211-215 (1984);

11 J.A. Thorpe: {\it Elementary Topics in Differential Geometry},
Springer-Verlag (1979);

12 R.M. Wald: {\it General Relativity}, Univ. of Chicago Press (1984).

\end{document}